\documentclass[letterpaper,12pt]{article}
\usepackage{cmap} 
\usepackage[english]{babel}
\usepackage[hidelinks,pdfpagelabels=false]{hyperref}

\usepackage{cite}
\usepackage{comment}
\usepackage{physics}
\usepackage{graphicx}
\usepackage{amssymb}
\newcommand{\be}{\begin{equation}}
\newcommand{\ee}{\end{equation}}
\newcommand{\bea}{\begin{eqnarray}}
\newcommand{\eea}{\end{eqnarray}}

\usepackage[ 
top=0.75in,
right=0.75in,
bottom=0.75in,
left=0.75in,
footskip=2em
]{geometry}
\usepackage{authblk}
\usepackage{fancyhdr}
\fancypagestyle{myfirstpage}{%
	\lhead{}
	\chead{}
	\rhead{INR-TH-2018-025}
	\lfoot{}
	\cfoot{}
	\rfoot{}
	
}

\usepackage{xifthen}

\def \ifempty#1{\def\temp{#1} \ifx\temp\empty }
\newcommand{\arXiv}[2]{arXiv:\href{https://arxiv.org/abs/#1}{#1\ifthenelse{\isempty{#2}}{}{~[#2]}}}
\newcommand{\doi}[1]{doi:\href{https://doi.org/#1}{#1}}
\newcommand{\email}[1]{\href{mailto:#1}{#1}}

\begin{document}
\title{Horndeski Genesis: strong coupling and absence thereof}

\author{Y. A. Ageeva\footnote{{\bf email:} \email{y.a.ageeva2604@gmail.com}} }
\author{O. A. Evseev\footnote{{\bf email:} \email{oa.evseev@physics.msu.ru}} }
\author{O. I. Melichev\footnote{{\bf email:} \email{oi.melichev@physics.msu.ru}} }
\author{V. A. Rubakov\footnote{{\bf email:} \email{rubakov@ms2.inr.ac.ru}} }
\affil{Institute for Nuclear Research of the Russian Academy of Sciences, 60th October Anniversary Prospect 7a, Moscow, 117312, Russia}
\affil{Department of Particle Physics and Cosmology, Faculty of Physics, M.~V.~Lomonosov Moscow State University, Vorobyovy Gory, 1-2, Moscow, 119991, Russia}

\date{September 30, 2018}
\maketitle
\thispagestyle{myfirstpage}
\vspace{-5pt}

\begin{abstract}
  
  We consider Genesis in the Horndeski theory as an alternative to or completion of the inflationary scenario. One of the options free of instabilities at all cosmological epochs is the one in which the early Genesis is naively plagued with strong coupling. We address this issue to see whether classical field theory description of the background evolution at this early stage is consistent, nevertheless. We argue that, indeed, despite the fact that the effective Plank mass tends to zero at early time asymptotics, the classical analysis is legitimate in a certain range of Lagrangian parameters.

\end{abstract}

\section{Introduction}
\label{intro}
Genesis  \cite{Creminelli:2010ba, Creminelli:2012my, Hinterbichler:2012fr, Hinterbichler:2012yn, Easson:2013bda, Nishi:2015pta, Nishi:2016wty} is a possible cosmological scenario in which the Universe starts its evolution from asymptotically
flat space-time at infinitely negative time. During the time evolution
the energy density, scale factor and Hubble rate grow. At some moment of time,
the Genesis regime is assumed to terminate,  and
conventional hot (or inflationary) epoch begins.


Genesis requires the violation of the Null Energy Condition (NEC) (for a review see, e.g., Ref.\cite{Rubakov:2014jja}). To violate
the NEC in a healthy way, one needs unusual matter. In a general non-canonical scalar field theory whose
Lagrangian depends on the scalar field $\phi$ and its first derivatives, the NEC can be violated. However, NEC-violating cosmological solutions are unstable because the curvature perturbation has either
wrong sign kinetic term \cite{ArmendarizPicon:1999rj, Garriga:1999vw} or
gradient instability or both. Healthy NEC violation can be obtained in
generalised Galileon/Horndeski
theory \cite{Horndeski:1974wa, Fairlie:1991qe, Fairlie:1992he, Fairlie:1992nb, Luty:2003vm, Nicolis:2004qq, Nicolis:2008in, Deffayet:2010zh, Deffayet:2010qz, Kobayashi:2010cm, Padilla:2012dx},  which is the most general scalar-tensor theory with second-order field equations.
Such a property is instrumental for avoiding  Ostrogradski instabilities, i.e. the ghost-like DOF that are usually associated with higher-order
time derivatives. In the original Genesis model and its versions, the
initial super-accelerating stage can occur without these instabilities \cite{Creminelli:2010ba, Creminelli:2012my, Hinterbichler:2012fr, Hinterbichler:2012yn, Easson:2013bda, Nishi:2015pta, Nishi:2016wty}. 

There is an issue in the Horndeski Genesis models, however. In most cases,
spatially flat Genesis solutions are plagued with gradient instabilities
occurring sooner or later in the cosmological
evolution~\cite{Libanov:2016kfc,Kobayashi:2016xpl,Kolevatov:2016ppi,Akama:2017jsa}.
This property has been formulated as the ``no-go theorem''.
One of the possible ways out is to consider models which are, at least
naively,
strongly coupled in the asymptotic past (and/or asymptotic future,
the case  that can be studied along the lines of this work)
\cite{Kobayashi:2016xpl,Ijjas:2016vtq,Nishi:2016ljg}. In these models, the
coefficients in quadratic action for perturbations about
the classical solution tend to zero as $t \to -\infty$, which, indeed,
implies that the strong coupling energy scale also tends to zero.

In this paper, we point out that this property does not necessarily
mean that one cannot use classical field theory for
describing the cosmological evolution at early times. Indeed, the
time scale of the classical evolution tends to infinity, and hence
its inverse, the
classical energy scale, tends to zero, as  $t \to -\infty$.
So, to see whether or not  the classical field theory treatment
is legitimate, one has to figure out the actual strong coupling energy
scale and compare it with the inverse time scale of the classical
background evolution. The classical analysis of the background is
consistent, provided that the former energy scale much exceeds the latter.
In this paper, we consider a simple class of Horndeski Genesis models
with the strong coupling at early times, and study scalar perturbations
in the asymptotics $t\to -\infty$.
We derive the conditions ensuring that the classical energy scale is
much lower than the strong coupling scale in the scalar sector.
We find that these conditions can indeed be satisfied in a certain range of
parameters in the Lagrangian, i.e.,
it is possible to avoid strong coupling regime
for Genesis stage at least as far as the scalar sector is concerned,
in the sense that the classical treatment
of the background evolution is consistent at early times.
We argue that tensor and tensor-scalar sectors may leave this
result unmodified.

This paper is organised as follows. In Section
\ref{general} we introduce the model and discuss its early-time
asymptotics that enables one to avoid the
%
no-go theorem
of Ref.~\cite{Kobayashi:2016xpl}.
In Section \ref{criteria} we
discuss strong coupling issue in detail and find a region of the
parameter space in which the classical description of
Genesis is legitimate despite the low strong coupling energy scale.
We conclude in Section \ref{conc}.

\section{Generalities}
\label{general}

\subsection{The model}

If one uses general relativity to describe gravity,
then an important characteristic is
the null energy condition (NEC) for the matter energy-momentum tensor
$T_{\mu\nu}$, which reads
$ T_{\mu\nu} k^{\mu}k^{\nu} \ge 0$ 
for every null vector $k^{\mu}$. Once the NEC holds in the cosmological context,
then (assuming flat spatial sections) it follows from the Einstein
equations that $dH/dt \le 0$, where $H$ is the Hubble parameter.
This implies that there is a singularity in the past of
the expanding universe. Therefore, one either modifies gravity
or violates the NEC to build non-singular cosmology. 

A candidate for NEC violating theory is the generalised
Galileon  scalar field coupled to
gravity~\cite{Creminelli:2010ba, Creminelli:2012my, Hinterbichler:2012fr, Hinterbichler:2012yn, Easson:2013bda, Nishi:2015pta, Nishi:2016wty}.
The most general form of Lagrangian which leads to the second-order field equations was obtained by G. Horndeski in \cite{Horndeski:1974wa}. It is sufficient for our purposes to consider
a subclass of Horndeski Lagrangians instead of the full one: 
\bea
    \cal L&=&G_2(\phi, X)-G_3(\phi, X)\Box \phi+ G_4(\phi)R, \nonumber\\
    X &=& -\frac{1}{2}g^{\mu\nu}\partial_{\mu}\phi\partial_{\nu}\phi,
    \label{Hor_L}
\eea
where $R$ is a Ricci scalar and $\Box \phi = g^{\mu\nu} \nabla_\mu \nabla_\nu \phi$. The metric signature is $(-,+,+,+)$.

Let us rewrite this Lagrangian \eqref{Hor_L} in terms of
ADM variables, to make contact with Ref.~\cite{Kobayashi:2016xpl}:
\begin{align}
    \mathcal{L} =  A_2 (t, N) + A_3 (t, N) K 
    +  A_4 (K^2 - K_{ij}^2) + B_4 (t, N) R^{(3)} \text{,}
    \label{adm_lagr}
\end{align}
where we use the unitary gauge in which $\phi = \phi(t)$, and $K_{ij}, R^{(3)}_{ij}$
are the extrinsic curvature and the Ricci tensor of the spatial slices, respectively.
There is one-to-one correspondence between the
variables
$\phi$ and $X$ in the covariant Lagrangian and time variable
$t$ and  lapse function $N$ in the ADM formalism.
The following expressions
convert one formalism to another \cite{Gleyzes:2013ooa, Gleyzes:2014dya, Fasiello:2014aqa}:
\begin{align}
    G_2 =& A_2 - 2XF_{\phi} \text{,} \label{ADM-trans2}\\
    G_3 =& - 2XF_X - F \text{,} \label{ADM-trans3}\\
    G_4 =& B_4 \text{,} \label{ADM-trans4}
\end{align}
where $F(\phi, X)$ is an auxiliary function, such that
\begin{equation}
    F_X = - \frac{A_3}{\left(2X\right)^{3/2}} - \frac{B_{4\phi}}{X} \text{,}
\end{equation}
and the following gauge is fixed with $Y_0=const$:
\begin{equation}
    e^{-\phi} = - \sqrt{2Y_0}t,\\ \nonumber
\end{equation}
so that
\begin{equation}
    e^{\phi}\sqrt{\frac{Y_0}{X}} = N.
    \label{gauge}
\end{equation}

\subsection{Avoiding the no-go theorem}

A subclass of Lagrangians in which the no-go theorem
can be avoided was given in Ref.~\cite{Kobayashi:2016xpl}:
\begin{align}
    &A_2 = M_{Pl}^4 f^{-2(\alpha +1) -\delta} a_2 (N) \text{,}\nonumber\\
    &A_3 = M_{Pl}^3 f^{-2\alpha -1 -\delta} a_3 (N) \text{,}\nonumber\\
    &A_4 = - B_4 = - M_{Pl} f^{-2\alpha} \text{,}
    \label{adm_func_lagr}
\end{align}
where $M_{Pl}$ is the
Planck mass, $\alpha$ and $\delta$ are constant parameters satisfying
\begin{equation}
    2\alpha > 1 + \delta \; ,  \hspace{5mm}  \delta > 0 \; , \label{Kob_cond}  
\end{equation}
and $f(t)$ is some function of time, which has the following asymptotics
as $t \rightarrow - \infty $
\begin{equation}
    f \approx -ct,  \hspace{5mm} c=\text{const}>0 \text{.}
\end{equation}
As a concrete example, we choose
\begin{align}
    &a_2(N) = -\frac{1}{N^2} + \frac{1}{3N^4} \text{,}\\
    &a_3(N) = \frac{1}{4N^3}\text{.}
    \label{smol_a}
\end{align}

The background  metric reads
\begin{equation}
    ds^2 = - N(t)^2 dt^2 + a(t)^2 dx_i dx^i,
\end{equation}
where $N$ is the lapse function (the same as in the
Lagrangian (\ref{Hor_L})). 
One derives the equations of motion for the
homogeneous background directly from the
variation of the background part of the Lagrangian \cite{Kobayashi:2015gga}
\begin{equation}
    \mathcal{L}^{(0)} = Na^3(A_2 +2 A_3H+6A_4H^2),
\end{equation}
and obtains
\begin{align}
    &(NA_2)_N+3NA_{3 \, N} H+6N^2(N^{-1}A_4)_N H^2 = 0, \\
    &A_2-6A_4H^2-\frac{1}{N}\frac{d}{dt}\left( A_3+4A_4H \right)=0, 
\end{align}
where the Hubble parameter is $H= \dot{a}/(Na)$  and subscript $N$ denotes the derivative upon lapse function $N$.
From these equations we find
an asymptotic solution at early times ($t \rightarrow - \infty$):
\begin{equation}
    H \approx \frac{\chi}{(-t)^{1+\delta}} \; ,
    \label{hub}
\end{equation}
\begin{equation}
    a \approx 1 + \frac{\chi}{\delta (-t)^\delta}, \quad N \approx 1\; ,
    \label{lapse_scale}
\end{equation}
where $\chi$ is the combination of the Lagrangian parameters
\be
\chi = \frac{\frac{2}{3}M_{Pl}^2+\frac{c}{4}\left(2\alpha+1+\delta\right)M_{Pl}}{4(2\alpha+1+\delta)c^{2+\delta}} \; .
\ee
An important feature of this solution is that
\begin{equation}
  B_4(t,N), ~A_4(t,N)
  \rightarrow 0 \quad \text{as} \quad t \rightarrow - \infty,
  \label{no-no-go1}
\end{equation}
and hence
\begin{equation}
  G_4(\phi,X) \rightarrow 0 \quad \text{as} \quad t \rightarrow - \infty \; .
  \label{no-no-go}
\end{equation}
On the one hand, these are  necessary conditions to avoid both 
ghost and gradient instabilities during subsequent evolution~\cite{Kobayashi:2016xpl}.
On the other hand, Eqs.~\eqref{no-no-go1} and \eqref{no-no-go}
signalise that the strong coupling energy scale in this
theory tends to zero as $t \to -\infty$. The purpose of this paper is
to see whether or not the latter feature spoils the classical
field theory description of the early time evolution, $t \to -\infty$.

\section{Strong coupling scale for perturbations
   versus classical scale}
\label{criteria}
We  now consider
the 
perturbations about the classical solution 
and, for technical reasons, study scalar perturbations only.
We comment on tensor and cross (tensor-tensor-scalar and scalar-scalar-tensor) sectors later on.
The perturbed metric for the
scalar sector has the following form
\begin{equation}
ds^2=-N^2 dt^2+\gamma_{ij}\left( dx^i+N^i dt\right)\left(dx^j+N^j dt\right),
\end{equation}
where 
\begin{align}
    N=1+\alpha \; , \;\;\;\; 
    N_i=\partial_i\beta\; , \;\;\;\; 
  \gamma_{ij}=a^2e^{2\zeta} \delta_{ij} \; ,
\end{align}
and $\alpha$, $\beta$, $\zeta$ are scalar perturbations.
Expanding the action up to the second order, one obtains
the following expression for the quadratic action in the unitary gauge
\begin{equation}
	\begin{aligned}
	S^{(2)}_{\alpha,\beta, \zeta} &= \int Ndt\text{ }ad^3x   \left[-3 g_\zeta \left(\frac{a}{N}\dot{\zeta}\right)^{2}+c_{\zeta}\left(\partial\zeta\right)^2 
  	  -  
          3a^2H^2 m_{\alpha} \alpha^2 +  2g_{\zeta}\partial\alpha\partial\zeta + 6\frac{a^2}{N}Hf_{\alpha}\alpha \dot{\zeta} 	\right. \\
  	& ~~~~~~~~~~~~~~~~~~~~~~~+\left. 
 2\frac{a}{N}g_{\zeta}\zeta\partial^2\beta - 2 aH f_{\alpha}\alpha\partial^2\beta  \right], \label{2act}
	\end{aligned}
\end{equation} 
where  
$\partial$ denotes spatial derivatives,
and
\begin{align}
   g_\zeta &= 2(B_4 + NB_{4N}), \nonumber \\
   c_\zeta &= 2B_4, \nonumber \\
   f_\alpha &=   2\left(\frac{NA_{3N}}{4H}+B_4 - NB_{4N}-N^2B_{4NN}\right), \nonumber \\
   m_\alpha &=  B_4 - NB_{4N}+2N^2B_{4NN} + N^3B_{4NNN}  \nonumber \\ &- \frac{1}{6H^2}\left(A_2+3NA_{2N}+N^2A_{2NN}\right) - \frac{1}{2H}\left(NA_{3N}+N^2A_{3NN}\right).
  \end{align}
The early-time asymptotics for the background solution (\ref{hub}) and (\ref{lapse_scale}) of the latter coefficients are
\begin{subequations}
  \label{aug17-18-vr1}
\begin{align}
    &g_{\zeta} \sim c_{\zeta} \sim (-t)^{-2\alpha}\; , \\
    &f_{\alpha} \sim (-t)^{-2\alpha}\; , \\ 
    &m_{\alpha} \sim -(-t)^{-2\alpha+\delta}\; .
\end{align}
\end{subequations}
The fields $\alpha$ and $\beta$ are constraint variables. One finds them
by solving the constraint equations and plugs them back into the action (\ref{2act}).
In this way one obtains 
%
the following expression for the unconstrained quadratic action:
\begin{equation}
	\begin{aligned}
	S^{(2)}_{\zeta} &= \int Ndt\text{ }ad^3x  \left(\frac{\epsilon_s}{c^2_s}\frac{a^2}{N^2}\dot{\zeta}^2-\epsilon_s(\partial \zeta)^2\right), \label{2act_no_alpha}
	\end{aligned}
\end{equation} 
where 
\begin{equation}
    \epsilon_s = \frac{1}{aN} \frac{d}{dt}\left(\frac{a g_\zeta^2}{H f_\alpha}   \right) - c_\zeta, \quad c_s^2 = \frac{\epsilon_s}{3g_\zeta}\left( 1 - \frac{g_\zeta m_\alpha}{f_\alpha^2} \right)^{-1} \; .
\end{equation}
The asymptotic behaviour of the functions $\epsilon_s$ and $c_s$
is found from \eqref{aug17-18-vr1}:
\begin{equation}
    \epsilon_s \sim (-t)^{-2\alpha+\delta}, \quad c_s^2 \sim (-t)^0 \; .
\end{equation}
Since $2\alpha - \delta > 1$, see \eqref{Kob_cond}, the overall
coefficient
$\epsilon_s$ tends to zero as $t \to -\infty$, signalling the low
strong coupling energy scale at early times.

To figure out the strong coupling scale in the scalar sector, we have
to go one step further and consider the cubic action.
We use the results presented in \cite{Gao:2011qe} for cubic action for all of the
scalar perturbations $\alpha$, $\beta$ and $\zeta$:

\begin{eqnarray}
    &  &S^{(3)}_{\zeta,\alpha,\beta} = \int Ndt~ ad^{3}x 
    \Big\{g_{\zeta}\Big[-9\frac{a^2}{N^2}\zeta\dot{\zeta}^{2}+2\frac{a}{N}\dot{\zeta}\Big(\zeta\partial^{2}\beta+\partial_{i}\zeta\partial^{i}\beta\Big)-\alpha(\partial_{i}\zeta)^{2}+(\partial_{i}\beta)^{2}\partial^{2}\zeta- \nonumber\\
    &  & -\frac{1}{2}\zeta\Big(4\alpha\partial^{2}\zeta-(\partial^{2}\beta)^{2}+(\partial_{i}\partial_{j}\beta)^{2}\Big)\Big]+  \nonumber\\
    &  & +c_{\zeta}\zeta(\partial_{i}\zeta)^{2}-9(aH)^2m_{\alpha}\alpha^{2}\zeta+2aHf_{\alpha}\alpha\Big(9\frac{a}{N}\zeta\dot{\zeta}-\zeta\partial^{2}\beta-\partial_{i}\zeta\partial^{i}\beta\Big) \nonumber\\
    &  & +\frac{\lambda_{1}}{aH}\Big[\frac{a^3}{N^3}\dot{\zeta}^{3}-\frac{a^2}{N^2}\dot{\zeta}^{2}\partial^{2}\beta+\frac{1}{2}\frac{a}{N}\dot{\zeta}\Big(4\alpha\partial^{2}\zeta+(\partial^{2}\beta)^{2}-(\partial_{i}\partial_{j}\beta)^{2}\Big)-\alpha\Big(\partial^{2}\zeta\partial^{2}\beta-\partial_{i}\partial_{j}\zeta\partial_{i}\partial_{j}\beta\Big)\Big] \nonumber\\
    &  & +\lambda_{2}\alpha\Big[3\frac{a^2}{N^2}\dot{\zeta}^{2}-2\frac{a}{N}\dot{\zeta}\partial^{2}\beta+\frac{1}{2}\Big((\partial^{2}\beta)^{2}-(\partial_{i}\partial_{j}\beta)^{2}\Big)\Big]- \nonumber\\
    &  &-\lambda_{3}aH\alpha^{2}\Big(3\frac{a}{N}\dot{\zeta}-\partial^{2}\beta\Big)-\lambda_{4}\alpha^{2}\partial^{2}\zeta+\frac{\lambda_{5}}{2}(aH)^{2}\alpha^{3}\Big\},
\end{eqnarray}
where $\lambda_1, \lambda_2, \lambda_3, \lambda_4, \lambda_5$ are the functions of $g_\zeta,c_{\zeta} f_\alpha, m_{\alpha}, A_2, A_3, A_4, H$ and we find their asymptotic behaviour as $t\to\infty$ for our model (\ref{adm_func_lagr}):
\begin{align}
\lambda_1 & = 0  ,\nonumber \\
\lambda_2 &\sim (-t)^{-2\alpha},\nonumber \\
\lambda_3 &\sim (-t)^{-2\alpha},\nonumber \\
\lambda_4 &\sim (-t)^{-2\alpha},\nonumber \\
\lambda_5 &\sim (-t)^{-2\alpha+\delta}.
\end{align}
We solve the constraints in terms of $\alpha$ and  $\beta$
and  obtain the following expression for unconstrained cubic action: 
\begin{align}
S^{(3)}_{\zeta} =  &\int Ndt\text{ }ad^3x \left\{ \Lambda_1 \left(\frac{a}{N}\dot{\zeta}\right)^3 + \Lambda_2 \left(\frac{a}{N}\dot{\zeta}\right)^2\zeta + \Lambda_3 \left(\frac{a}{N}\dot{\zeta}\right)^2 \partial^2 \zeta  +
  \Lambda_4 \left(\frac{a}{N}\dot{\zeta}\right)\zeta \partial^2 \zeta\right. \nonumber \\
  &  + \Lambda_5 \left(\frac{a}{N}\dot{\zeta}\right) \left(\partial_i \zeta \right)^2 
+ 
 +\Lambda_6 \zeta \left(\partial_i \zeta \right)^2 + \Lambda_7 \left(\frac{a}{N}\dot{\zeta}\right) \left(\partial^2 \zeta \right)^2 + \Lambda_8 \zeta \left(\partial^2 \zeta \right)^2
 \nonumber \\
&  + \Lambda_9 \partial^2 \zeta \left(\partial_i \zeta \right)^2 + \Lambda_{10} \left(\frac{a}{N}\dot{\zeta}\right) \left(\partial_i \partial_j \zeta \right)^2 + \Lambda_{11} \zeta \left(\partial_i \partial_j \zeta \right)^2 
  + \Lambda_{12} \left(\frac{a}{N}\dot{\zeta}\right) \partial_i \zeta \partial^i \psi \nonumber \\
  &+ \Lambda_{13} \partial^2 \zeta \partial_i \zeta \partial^i \psi + \Lambda_{14} \partial^2 \zeta \left( \partial_i \psi \right)^2 
 + \Lambda_{15} \left(\frac{a}{N}\dot{\zeta}\right) \left(\partial_i \partial_j \psi \right)^2 + \Lambda_{16} \zeta \left(\partial_i \partial_j \psi \right)^2 + \nonumber \\
 & + \left.
\Lambda_{17} \left(\frac{a}{N}\dot{\zeta}\right) \partial_i \partial_j \zeta \partial^i \partial^j \psi + \Lambda_{18} \zeta \partial_i \partial_j \zeta \partial^i \partial^j \psi \right\}, \label{S3}
\end{align}
where $\psi = \partial^{-2}(a\dot{\zeta}/N)$; $\Lambda_1 ... \Lambda_{18}$ are
functions of $g_\zeta,c_{\zeta} f_\alpha, m_{\alpha}, A_2, A_3, A_4, H$, and hence
of time
$t$. All of them have power-law behaviour at early times
$t\to-\infty$,
\be
\Lambda_i \sim (-t)^{x_i},
\ee
where $x_i$ are combinations of $\alpha$ and $\delta$.

For power-counting purposes,
every term $\mathbb{L}_i$ in the cubic Lagrangian  
 ($i = \overline{1,18}$) can be schematically written as follows
\be
    \label{term_lagr}
    \mathbb{L}_i = \Lambda_i\cdot  \zeta^3 \cdot (\partial_t)^{a_i} \cdot
    ( \partial )^{b_i}
    \; ,
    \ee
where $a_i$ and $b_i$ are the
numbers of temporal and spatial derivatives, respectively.
    
In our dimensional analysis below we naturally use the canonically
normalised field $\pi$ instead of $\zeta$. Since $a(t)$, $N(t)$ and
 $c_s^2 (t)$ 
tend to constants as  $t \to -\infty$, the canonically normalised field is
(modulo a constant factor)
\begin{equation}
  \pi = \sqrt{{\epsilon_s}} \zeta \propto |t|^{-\alpha +\delta/2} \zeta 
  \; . 
        \label{can}
\end{equation}
The fact that the coefficient here tends to zero  as  $t \to -\infty$
is crucial for what follows.

    In terms of the canonically normalised field $\pi$ one rewrites (\ref{term_lagr}) as follows:
\be
    \label{a_fin}
    \mathbb{L}_i = \tilde{\Lambda}_i\cdot  \pi^3 \cdot (\partial_t)^{a_i} \cdot
    ( \partial )^{b_i}  
    \ee
    where
    \be
    \tilde{\Lambda}_i = \Lambda_i\epsilon_s^{-3/2} =\Lambda_i |t|^{-\frac{3}{2}(\delta - 2\alpha)}
    \sim  |t|^{x_i-\frac{3}{2}(\delta - 2\alpha)} \; .
\ee
    By naive dimensional analysis (dimension of $\Lambda_i$ is $[\Lambda_i] = 4-a-b$ and $[\epsilon_s] = 2$) we immediately find
    that the strong coupling energy scale associated with the term
    $\mathbb{L}_i$ is
    \be
    E^{(i)}_{strong} \sim \tilde{\Lambda}_i^{- \frac{1}{a_i + b_i -1}}
    \sim |t|^{-\frac{x_i + 3\alpha - 3\delta/2}{a_i + b_i -1}} \; .
    \ee
    On the other hand, the inverse time scale of classical evolution
    is
    \be
    E_{class} \sim  \frac{\dot{H}}{H} \sim |t|^{-1} \; .
    \ee
    Thus, the condition for legitimacy of the classical treatment
    of the early evolution, $ E_{class} \ll  E^{(i)}_{strong}$ for all $i$
    reads
\be
x_i+3\alpha-\frac{3}{2}\delta < a_i+b_i-1 \; , \;\;\;\;\; i = \overline{1,18}\; .
\ee
Clearly, the most dangerous terms are those with the smallest combination
$a_i + b_i - x_i$. By inspecting the behaviour of $\Lambda_i$ one finds
that this combination is the smallest for $i=1$ (given the constraints~\eqref{Kob_cond}), when
\be
\Lambda_1 \sim (-t)^{1-2\alpha +3\delta}\; , \; \; \; a_1=3\; ,
\;\;\; b_1=0 \; , \;\;\;\; a_1+b_1-x_1 = 2 + 2\alpha - 3 \delta 
\ee
(as an example, the next term has  $\Lambda_2 \sim (-t)^{-2\alpha+2\delta}$,
$a_2 = 2$, $b_2=0$ and  $a_2+b_2-x_2 = 2 + 2\alpha - 2 \delta$; recall that
$\delta >0$).
Thus, 
the strong coupling regime can be avoided for
$2\alpha < 2- 3\delta$, which together with \eqref{Kob_cond} gives
\begin{equation}
0 < \delta < \frac{1}{4}, 
\qquad 2 - 3\delta > 2\alpha > 1 + \delta. 
\end{equation}

\begin{figure}[h]
\centering
\includegraphics[width=14cm]{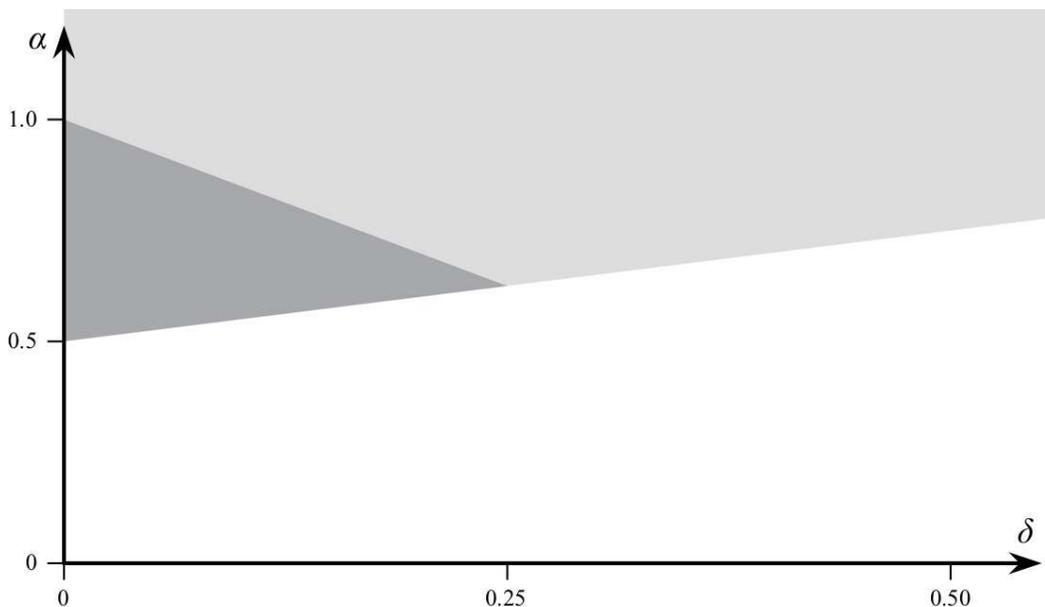}
\caption{Two grey regions together correspond to the space of the Lagrangian parameters. The light grey region
  corresponds to the Lagrangian parameters which yield Minkowski
  space-time at $t\to-\infty$ and avoid the no-go theorem of
  Ref.~\cite{Kobayashi:2016xpl}.  The dark grey area
  shows the allowed range of the
  Lagrangian parameters satisfying "no strong coupling" criterion.}
\label{possible_param}
\end{figure}

We conclude that the
strong coupling regime is
avoided (at least as far as the scalar perturbations are concerned), in
the sense that the evolution remains classical at early times,
provided  one chooses the
Lagrangian parameters $\alpha$ and $\delta$ in the dark grey allowed
region shown in Fig.~\ref{possible_param}.

To get more confidence in the classical field theory
treatment avoiding the strong coupling problem, one has to
analyse tensor, tensor-tensor-scalar
and scalar-scalar-tensor sectors of perturbations.
It is likely, though, that 
they  give weaker constraints than those
presented above.

\section{Summary} 
\label{conc}
We have studied the non-singular Genesis scenario
in the framework of the Horndeski theory, which is capable
of avoiding the gradient instability at the expense of potential
strong coupling problem. The model of Ref.~\cite{Kobayashi:2016xpl}
has been used as an example that gives explicit asymptotic solutions at early times.
We have seen that with an appropriate choice of parameters,
these solutions are actually away from the strong coupling
regime inferred from the study of scalar perturbations.
This opens up a possibility that the Universe starts up with
very low quantum gravity energy scale (effective Planck mass
asymptotically vanishes as $t \to -\infty$), and yet its classical
evolution is so slow that the classical field theory description
remains valid.

Even though our analysis has given a promising outcome, it is
certainly incomplete. First, we still have to study
tensor perturbations and their cubic self-interactions and interactions with
scalar perturbations. Second, 
there is no guarantee that the fourth
and higher order interactions give strong coupling energy
scales higher or equal to the ones we have found by studying the
cubic interactions. We hope to turn to these issues in future.

\section*{Acknowledgments} 
We would like to thank  E.~Babichev, P. ~Creminelli,  S.~Mironov, A.~Vikman and
V.~Volkova for numerous
helpful discussions. This work has been
supported by Russian Science Foundation grant 14-22-00161.

\end{document}